# Understanding People's Needs in Viewing Diverse Social Opinions about Controversial Topics

Hayeong Song*
Georgia Institute of Technology

Zhengyang Qi†
Carnegie Mellon University

John Stasko‡
Georgia Institute of Technology

Diyi Yang§
Stanford University

**ABSTRACT**

Social media (i.e., Reddit) users are overloaded with people's opinions when viewing discourses about divisive topics. Traditional user interfaces in such media present those opinions in a linear structure, which can limit users in viewing diverse social opinions at scale. Prior work has recognized this limitation, that the linear structure can reinforce biases, where a certain point of view becomes widespread simply because many viewers seem to believe it. This limitation can make it difficult for users to have a truly conversational mode of mediated discussion. Thus, when designing a user interface for viewing people's opinions, we should consider ways to mitigate selective exposure to information and polarization of opinions. We conducted a needs-finding study with 11 Reddit users, who follow climate change threads and make posts and comments regularly. In the study, we aimed to understand key limitations in people viewing online controversial discourses and to extract design implications to address these problems. Our findings discuss potential future directions to address these problems.

**Index Terms:** Human-centered computing—Visualization—Online communities—Discourse analysis; Human-centered computing—Visualization—Formative study

## 1 INTRODUCTION

Internet forums such as Reddit and Twitter provide millions of people access to online conversations and discussions. But a few limitations exist in viewing online discourses such as their large-scale, polarization, and their linear structure. For example, online discussions on Reddit are represented in a linear structure that does not provide an overview of the comments, which makes it difficult for people to capture a high-level synthesis of the discussions. Prior works on discourse architecture studies have recognized this limitation of linear online comment structure (list) [27, 31] in viewing data at scale. This limitation can make it difficult for users to navigate to insightful comments, which reduces learning effects between discussants by viewing others' opinions [31]. Users also suffer from lacking moderation between discussants [27] which makes it difficult for users to have a truly conversational mode of mediated discussion.

Along those lines for a controversial topic, the incoherent structure can reinforce biases, "cyberpolarization", in which a certain point of view becomes widespread, simply because many viewers seem to believe it [5]. Thus, when we design a user interface for viewing people's opinions we should consider ways to mitigate selective exposure to information and polarization of opinions [11, 19, 23, 24]. To address these problems, prior research showed that system designs and structures can assist people to view opposing views and highlight different aspects of the issues that can mitigate the problem of selective exposures [19, 25]. For example, Considerit [18] is a platform that helps users to deliberate on diverse perspectives about difficult decisions such as U.S. state elections. The system provides a workflow to reflect the perspectives of others by framing interactions around pro/con points that participants create, adopt, and share.

Up to this point, prior research has focused on assisting people to view online conversations at scale, such as providing summarization workflow [33], showing an overview of the discussion space, or showing evolving topics [15, 16]. But less work has been done on assisting users to view online discourses about a controversial topic such as climate change, to mitigate people's opinion polarization. To address these problems, we conducted a needs-findings study with 11 Reddit users who follow climate change threads and actively make posts and comments on the climate change subreddit. We conducted semi-structured interviews and surveys. The interview themes were the difficulties people face in viewing controversial online discourses and what people look for in viewing discussion space. Based on these findings we discuss design implications to address these problems.

We provide the following contributions to the visualization research community:

- Needs-finding study results and their analysis
- Design implications derived from the study analysis

## 2 RELATED WORK

We build our research on two major areas of related work (1) discourse analysis and (2) visualizing diverse perspectives.

### 2.1 Discourse analysis

In this section, we describe challenges that people encounter while viewing and analyzing online discourses, and we discuss prior approaches that aimed to alleviate these issues. Some of the challenges that exist in viewing online discourses are large-scale, polarization, and linear structure. Dahlgren [8] acknowledged this limitation of a linear discourse architecture, where inherent bias is caused by the fragmentation of participants. Here, fragmentation of participants means that people lean toward viewing only like-minded views, which can deter mutual understanding between discussants about a polarized topic. Thus, political discussion and public opinions are susceptible to inherent bias [4].

To alleviate this issue, visualization tools such as Opinion space [11] were designed in a way that assists users to reflect on other people's opinions [12]. For example, in the system, people's opinions are presented in a two-dimensional scatter plot view. Then each data point represents a comment and the proximity of data points indicates the similarity of opinions. If two people's opinions are similar, their distance is close to each other and vice versa. Also, highly-rated opinions are highlighted in green whereas low-rated comments are highlighted in red. With this view, people can reflect on other people's opinions asynchronously by browsing a range of perspectives.

*e-mail: hsong300@gatech.edu
†e-mail: zqi2@andrew.cmu.edu
‡e-mail: stasko@cc.gatech.edu
§e-mail: diyiy@cs.stanford.edu

## 2.2 Visualizing diverse perspectives

To visually represent diverse perspectives, prior work tightly coupled visualizations with natural language processing techniques to process and extract data such as topics, public sentiments, and people's opinions [7, 20–22, 29, 30]. For example, some of the prior work leveraged different ways of supporting interactive topic model views to help people to capture high-level synthesis of the discussion space. These include supporting interactive views that show a hierarchy of topics [10], showings trend (over time) of topics being discussed [6], or dynamically presenting merging and splitting topics [9].

Some of the existing visualization tools, such as MulticonVis [16], a text analytics system for exploring a collection of online conversations, use topic modeling and sentiment analysis with information visualizations to support users in viewing large-scale blog conversations. To help users to explore conversations and give more context about the conversations, the system supports a view that shows sentiment distribution in a stacked graph next to a conversation. Gao's intelligent system [13] exposes users to view diverse perspectives by showing people's sentiments about controversial topics such as U.S. presidential candidates. The goal of the system is to expose users to other people's opinions and appreciate people with different stances in online forums. This is done by using interactive visualization and categorizing original posts based on crowd workers' reactions and emotions from different stances.

Although their approach took steps to encourage users to view diverse perspectives and public reflection on people's opinions, it is more curated to the scenario that they were using (i.e., confined to showing emotions [13]). This can make it difficult to generalize for designing a system that helps people to view online discourses on a divisive topic because of online forums (e.g., Reddit).

## 3 DISCUSSION SPACE

In this section, we discuss our approach to understanding the online discussion space, which studies people's posting and commenting patterns for three controversial topics. To understand the online discussion space about a polarized topic on Reddit and understand the relevant data, we explained people's posting patterns (e.g., how many posts were made). For our analysis, we analyzed three different topics: climate change, gun control, and abortion. We selected these topics based on Beel et al's work on divisive online topics [3].

**Dataset.** For the analysis, we collected data for the time frame 01/01/2022 to 08/30/2022 using PushShift API. For each topic, we collected data from subreddits r/climatechange, r/guncontrol, and r/abortion. We selected these subreddits because each was one of the threads where people posted regularly and had active discussions. Each subreddit thread had about r/climatechange: 61.3k, r/guncontrol: 11k, r/abortion: 33.7k number of members.

**Analysis.** We conducted this analysis to understand how many posts and comments are made during a certain duration of time. We also aimed to understand, how many posts or comments users make on average, and the maximum number of posts or comments a unique user makes. We also wanted to observe if there are differences in posting and commenting patterns across topics. Here, a post starts a new topic and a comment is a reply to a post.

**Post.** For posting patterns on topic *climate change*, 2913 posts were made over the period of time, and each user made 0.5 posts on average (see Table 1). The maximum number of posts a unique user made was 112. For topic *gun control*, 897 posts were made over the period of time, and each user made 0.6 posts on average. The maximum number of posts a unique user made was 44. For topic *abortion*, 8102 posts were made over the period of time, and each user made 0.8 posts on average. The maximum number of posts a unique user made was 33.

**Comment.** For commenting patterns on the topic *climate change*, 27811 posts were made over the period of time, and each user made 5 posts on average (see Table 2). The maximum number of posts a unique user made was 2232. For the topic *gun control*, 10325 posts were made over the period of time, and each user made 7 posts on average. The maximum number of posts a unique user made was 3655. For the topic *abortion*, 57437 posts were made over the period of time, and each user made 6 posts on average. The maximum number of posts a unique user made was 8303.

| Post | Climate change | Gun control | Abortion |
|---|---|---|---|
| Count | 2913 | 897 | 8102 |
| Per user | M=0.5,SD=2.1 | M=0.6,SD=2 | M=0.8,SD=1.2 |
| Max | 112 | 44 | 33 |

Table 1: Summary of the people's posting patterns for three divisive online topics on Reddit. M refers to mean and SD refers to standard deviation.

| Comment | Climate change | Gun control | Abortion |
|---|---|---|---|
| Count | 27811 | 10325 | 57437 |
| Per user | M=5,SD=37 | M=7,SD=105 | M=6,SD=110 |
| Max | 2232 | 3655 | 8303 |

Table 2: Summary of the people's commenting patterns for three divisive online topics on Reddit. M refers to mean and SD refers to standard deviation.

In our observations, we did not see a large difference in posting or commenting patterns between the topics *climate change* and *gun control*. But we noticed that post and comment numbers on topic *abortion* were higher compared to the other topics. We speculate that posting and commenting patterns on abortion were more active compared to other topics due to recent discussions about abortion legislation (US abortion debate).

## 4 FORMATIVE STUDY

After learning about people's posting and commenting patterns, we wanted to better understand what are the difficulties that people face in following online discourses and what argumentation people desired to better understand the discussion space. To discover this, we designed and conducted a needs finding [32] study with active Reddit users. For the study, we used the scenario of following climate change discussions because it was a topic that is controversial and subject to polarization. We conducted a formative study with 11 participants who follow climate change discussion threads on Reddit. We conducted surveys and semi-structured interviews that asked about their experience in following climate change discussion threads on Reddit, what they look for in the discussion space, and how visualization can potentially help.

### 4.1 Study design

**Participants.** We recruited 11 active Reddit users (8 male, 3 female), with backgrounds that included engineering, research, and development. Here we define active users as participants who follow climate change threads on Reddit and regularly visit and make posts and comments on the thread. In a Likert scale survey, they reported that on average they visit climate change threads on Reddit and make comments multiple times a week (M = 9.73 and SD = 0.62) (1 - never, 10 - multiple times a week). All of the participants reported being very familiar with current climate change discourses. We recruited participants from a local university via word-of-mouth and flyers.

**Procedure.** Each session was about 30-40 minutes long. The study was conducted remotely via Microsoft Teams. When a participant joined a study session, we asked the participant to fill out a survey. The survey included questions such as what were the difficulties in following climate change discussions with the current interface, what could be improved in the current interface, and what

7

climate-related topics they are interested in understanding. We also collected demographic information in the survey.

After participants filled out the survey, we transitioned into a semi-structured interview session. In the interview, we asked questions that had similar themes to the survey, which asked about their experience in following climate change discussions on Reddit, what they look for in the discussion space, and improvement points they desired in making sense of climate change discussions. We also inquired about their experience with visualization and potential points where they see visualization could help them make sense of climate change discussions. In this session, we often followed up with questions to capture more details and insights to clarify some of the participants' points.

**Data analysis.** We conducted an exploratory qualitative analysis of the recordings using an open-coding approach. We looked for themes, such as when participants encountered difficulties in following climate change discussion threads on Reddit, what they desired to make sense of the discussion space, and improvement points of the current interface that could make navigation of the discussion space easier. We also looked into when and how visualization could potentially help people make sense of climate change discussions. One researcher coded all eleven interviews and a second researcher independently coded all the interviews for validation. Coders noted similar themes in the interviews and were able to further concretize similarities and differences.

### 4.2 Results & Key Findings

In this section, we discuss our study results and key findings. We discuss specific topics that interested participants, some of the difficulties that participants faced, and what they desired to better understand the discussion space. Our major emergent finding was that participants desired an expansion of information about authors in the discussion thread and desired ways to view diverse perspectives more easily. They also desired visual hints of the climate change discussion thread that can help with the navigation of the discussion space. We coded our participants as P1-11. We call out individual support for each point based on these participant codes.

#### 4.2.1 Current Practices

In this section, we describe participants' current practices in viewing online discourses on Reddit. We discuss the topics that they were most interested in the discussion space and the difficulties that participants faced in following online discourses.

- **Interested topics.** 9 out of 11 participants reported their specific interest in topics such as climate consequences, climate policies, and solutions. For example, these participants were interested in climate consequences and their effects on human lives. They also wanted to view climate solutions and policies to reduce climate consequences. The other two reported that they were interested in viewing event information (e.g., hurricanes) and calling for climate action. P3 stated that *"I am mainly interested in understanding climate policies and climate solutions, such as climate policies across countries."*

- **Difficulties.** All of our participants pointed out a current limitation in the list like structure in viewing online discourses at scale [27, 31], such as when threads get extremely long it becomes difficult to find insightful comments quickly and view diverse perspectives. Here, we refer to insightful comments that can provide an in-depth understanding of the topic being discussed. P3 stated *"It can be difficult to find insightful comments sometimes when the thread gets too long. The comments easily pile up and can be hard to navigate those comments.* Also, they acknowledged the problem of selective exposure to opinions, when the discussion space seemed to have mostly like-minded views [19, 25]. P10 stated *"I remember when I was trying to understand the topic, I wished to see an opposing view which was so difficult to do because the thread was dominated by so many people who had like-minded opinions."*

#### 4.2.2 Improvement points

In this section, we discuss what participants seek to help them better understand the discussion space.

- **Expansion of information.** 6 out of 11 participants desired expansion of information about authors' background (e.g., country of residence (location), the threads they follow on Reddit, and their political stance) [P1, P4, P5-7, P11]. They desired this information because it would give them more context when navigating and following the discussion thread. For example, they were curious about understanding public sentiment towards a climate policy per country or based on people's political stance (e.g., Republican vs Democrat). P11 stated that *"If it annotates the comment based on where the person is located, that would help me understand the context of the discussion based on their location as different countries have different climate policies."*

- **Desired visual hints of the discussion structure.** 10 out of 11 participants desired visual hints (e.g., annotation with labels) of the discussion structure to capture a high level-synthesis of the discussion space [P1-10]. They desired this information because it was often difficult to follow the list-like structure because of the scale and because the topic changes spontaneously. Thus, they desired ways to navigate the discussion space more effectively such as by being able to find insightful comments quickly. P3 stated *"Actually, I wish we can have some kind of filter that automatically filters important opinions and important comments, I would like to find that quickly with annotations or highlights [visual hints]. Maybe the comment that got the most reactions."* P8 stated that *"Actually, I wished that there is some way to see an overview [discussion structure] of the discussion because the topic changes so spontaneously it's hard to follow sometimes."*

  Also, they desired to see opinions in groups, such as supporting or against, to see dominating opinions about a certain discussion topic. P1 stated that *"So if it somehow [visual hints] shows an overview of some of the [pro/con] discussion points that would be helpful."* P6 stated *"If I want to make comparisons with other people's opinions, I wanted to see them in clusters. If I am against [this policy] I wanted to see people's opinions who support them, but see them in groups with some indication [visual hint]"*.

- **View diverse perspectives.** 7 out of 11 participants desired ways to view diverse perspectives and desired to have a balanced view of the topic [P1, P3-7, P9-11]. Participants stated that they particularly like the community vibe that Reddit provides compared to other social media platforms such as Twitter and Facebook. They stated that with the community vibe, they can view a range of people's different opinions and learn from each other. Thus, they desired ways to see other people's opinions more efficiently.

  For example, P9 stated *"In my opinion, [when I navigate the discussions space] I look for diverse opinions, not just my opinion. I look for ways to see diverse opinions, so I can learn more so and have more ideas about others' perspectives."* P5 also stated *"When it comes to reading it, I love to see other people's opinions about climate change, so I'll compare it with my own to get more information and have a balanced view"*.

## 5 DESIGN IMPLICATIONS

We further coded the interviews to extract design implications for developing an augmented user interface that can assist people to follow climate change discussions. Based on our key findings and design implications extracted from the formative study, we propose the following design goals. Our goals leverage insights from prior work and study findings. We refer to the four design goals as G1-G4.

- **G1. Provide high-level synthesis.** We propose to visually represent a structure (e.g., reply structure) of the discussion space to provide an overview. We propose this goal to design an "overview + detail" type of interface to assist people to view the discussion space at scale. This way users can drill down to specifics after getting an overview. For example, an interface can visualize the distribution of both sides (pro/con) of the debated issue to provide an "overview" of the discussion and users can retrieve individual opinion that falls under that category ("detail").

- **G2. Assist in viewing diverse perspectives.** Online discourses are susceptible to "cyberpolarization", where some of the opinions get attention simply because they seem to be prevalent in the discussion space. This takes away an opportunity of learning from other participants by viewing their opinions. This goal focuses on leveraging ways to assist people in viewing a range of diverse perspectives including minority opinions.

- **G3. Provide context in the discussion space.** Participants desired expansions of information and contextualized information about users, such as an author's political stance or reputation of the author in the community. We seek methods to provide that information in a practical way in the system.

- **G4. Allow easier discovery of insightful comments.** We seek ways to help people navigate to insightful comments quickly. We propose to provide visual hints (visual guidance), with annotation on the view to help people find those comments, such as providing ways to filter or recommend comments based on user scores (e.g., upvotes) by the reputation of the author in the discussion community. We propose this design goal to help users to move through the content easily by providing navigational affordances [14, 28] in the discussion space.

## 6 DISCUSSION

Our formative study revealed that users desired:
- Expansion of information
- Visual hints of the discussion structure
- Assistance in viewing diverse perspectives

Based on these findings, we wanted to further investigate designing augmented reading interfaces to specifically assist readers in viewing online discourses about a divisive topic.

### 6.1 Expansion of information

Some of the prior work focused on presenting metadata such as an author and when the comment or opinion was created. While these can be informative, presenting metadata alone could be insufficient to engage the public effectively in discussions. For example, users may want to know more about authors, such as their background, political stance, and their interests. By providing expansion of information it can help people draw attention to political aspects or questions depending on who is performing that interpretation. To provide such capabilities, we discuss what information can be provided additionally in the discussion space.

**Stances (Opposing vs Supporting).** In a visual analytics system, people's comments can be annotated based on their stances so that users can see comments in clusters and infer the ratio of support and against opinion on a certain topic. For example, Procon.org [1], a non-profit organization, presents this kind of view, where perspectives from both proponents and opponents are shown in a table with color-codings about controversial issues. To extract people's stances, one could potentially use the method described by Hosseinia [17].

**Political stances (Republican vs Democrat).** Participants desired information about users' political stances to contextualize their comments and posts when viewing the discussion. Thus, one could consider providing such information in the system. To extract people's political stance, one could potentially use method and heuristics by Rajadesingan's [26] work.

### 6.2 Author's background and reputation

Participants desired information about people's backgrounds and interests to unpack the reasoning behind their opinions. For example, providing the location of the authors will help users get more context in viewing climate policies in a discussion space because different countries have different legislation. Providing subreddits that users follow can be useful in knowing what other users are interested in to get recommendations.

Another recommendation is to suggest comments or posts worth checking based on the author's reputation in the community (e.g. karma score, upvote score), as they are known for making higher-quality posts and comments. This can help users to navigate to insightful comments and post quickly. This can potentially help users to see a range of diverse perspectives and be informed about other people's opinions.

This information can also impact people's openness to new opinions. Having access to authors' backgrounds and reputations can impact how people perceive other people's opinions (e.g., trustworthiness or quality). For example, Reddit supports a feature where users can request a user flair (e.g., types of labels such as Ph.D. Climate Science) for their accounts. Then these labels show next to their post or comments. This additional information can put more weight on their opinions compared to others who don't have those titles. We envision that providing an additional layer of information can do a similar job, which can encourage users to read and be open to accepting new opinions.

### 6.3 Aggregating people's opinions

Although prior work (e.g., multiConvis [16], Wikum [33]) showed aggregations of data such as an overview, clusters of opinions, and sentiment, may simply lead to showing dominant opinions, where this can lead to marginalizing some of the less dominant opinions. But this might be a type of information that a policymaker, a potential user of a visualization system, would want to know to well represent unpopular opinions or minority groups. Thus, when aggregating data we should think about granularity for visualization to find the right balance so that minority opinions do not get missed. This can translate to system designs to leverage the interplay between visualization decisions and how it can lead to people's interpretation of data for controversial or political topics [2].

## 7 CONCLUSION

The goal of our work has been to understand user needs and extract design implications for a system that can assist people in viewing online discourses about a divisive topic. To this end, we conducted a needs-finding study with 11 Reddit users who follow climate change threads. Our findings showed that participants desired expansion of information, and visual assistance to view diverse perspectives and capture high-level synthesis of the discussion space. Based on these identifications of user needs we discuss design goals that can potentially be used in other system design that deals with divisive online topics.